\documentclass[jcp,aip,amsmath,amssymb,reprint,USenglish,nofootinbib,floatfix]{revtex4-2}

\usepackage{graphicx}
\usepackage{dcolumn}
\usepackage{floatflt,epsfig} 
\usepackage{mathtools}
\usepackage{color}
\usepackage{braket}
\usepackage{hyperref}
\usepackage[normalem]{ulem}
\usepackage{blindtext}
\usepackage{lipsum}  
\usepackage{amsthm}
\usepackage{tabularx}
\usepackage{bm}
\usepackage{dsfont}
\usepackage{bbold}
\usepackage{ulem}
\usepackage{mwe,tikz}
\usepackage[percent]{overpic}
\usepackage{todonotes}
\usepackage{algorithm}
\usepackage{algpseudocode}
\usepackage{comment}

\usepackage{xr}
\makeatletter
\newcommand*{\addFileDependency}[1]{
  \typeout{(#1)}
  \@addtofilelist{#1}
  \IfFileExists{#1}{}{\typeout{No file #1.}}
}
\makeatother
\newcommand*{\myexternaldocument}[1]{
    \externaldocument{#1}
    \addFileDependency{#1.tex}
    \addFileDependency{#1.aux}
}
\myexternaldocument{si}

\begin{document}

\title{Sampling Free Energy Landscapes of Ionic Colloidal Crystal Systems using Machine-Learned Proxy Collective Variables}

  \author{Michael S. Chen}
  \email{michael.chen@uky.edu}
    \affiliation{Department of Chemistry, New York University, New York, NY 10003 USA}
  \affiliation{Simons Center for Computational Physical Chemistry, New York University, New York, NY 10003 USA}
  \affiliation{Department of Chemical and Materials Engineering, University of Kentucky, Lexington, KY 40506 USA}
  \affiliation{Department of Computer Science, University of Kentucky, Lexington, KY 40506 USA}

  \author{Stefano Sacanna}
  \affiliation{Department of Chemistry, New York University, New York, NY 10003 USA}

  \author{Glen M. Hocky}
  \email{hockyg@nyu.edu}
    \affiliation{Department of Chemistry, New York University, New York, NY 10003 USA}
  \affiliation{Simons Center for Computational Physical Chemistry, New York University, New York, NY 10003 USA}

\date{\today}

\begin{abstract}
 Charged colloids coated with a polymer brush can be designed to preferentially self-assemble into different crystal structures by varying easy-to-tune experimental conditions. For a given set of conditions, we have observed in experiments and simulations a distribution of thermodynamically (meta)stable self-assembled crystal structures. Properly quantifying the free energy landscape of these colloidal systems is essential for rationally choosing conditions to preferentially target particular crystal structures. For some of the structures we have formed, standard crystalline order parameters are not able to differentiate between crystals or between crystals and amorphous aggregates. We show that local environment similarity descriptors are able to distinguish the relevant metastable states, but are too expensive for use in biased MD simulations.  Here, we adopt an approach from machine-learned interaction potentials showing that SE(3)-equivariant transformer networks can serve as an efficient-to-evaluate machine-learned proxy. As a result, we can compute the relative free energies of accessible colloidal structures as a function of different experimentally-relevant physical knobs that can steer our system between two observed crystal types. As an example application, we then show how changing surface potentials of positive and negative colloids while maintaining the same attractive energy can shift which crystal structure is favored. 
\end{abstract}

\maketitle

\section{Introduction}
\label{sec:intro}
Colloidal particles of nano- to micrometer diameters have garnered significant interest due to the tunability of their interactions and, consequently, the ability to steer their self-assembly into a variety of extended crystalline structures including those presenting with desirable mechanical\cite{Jansen2023nanocrystal,Michelson2023high,Li2023ultrastrong,Qian2025nanoscale}, optical\cite{Pattabhiraman2017novel,He2020colloidal,gales2025crystallization}, and electronic properties\cite{Colvin1994led,Dabbousi1995electroluminescence}.
A wide array of colloidal building blocks have been synthesized for different design challenges ranging from spherical colloidal particles of varying compositions to those that are anisotropic in shape or possess programmable interactions\cite{Murray2000synthesis,Glotzer2007anisotropy,Li2011colloidal,Sacanna2011shape,Du2011anisotropic,Boles2016self,Zhou2025engineering,Hueckel2021total,Leunissen2005ionic}.
Among these, polymer-attenuated Coulombic self-assembly (PACS)\cite{Hueckel2020ionic} is a powerful platform whereby oppositely charged colloidal particles interact with one another via electrostatic forces that are attenuated at close range by interactions between the polymer brushes coating their respective surfaces.
We have demonstrated that this versatile approach allows us to tune the interparticle interactions of these binary ionic colloidal systems to promote the preferential self-assembly of a menagerie of different crystal structures and different nucleation dynamics\cite{Hueckel2020ionic,Zang2024structures,Zang2025direct,vanKesteren2026light}.

For a given set of PACS conditions, various thermodynamically stable and kinetically accessible crystal polymorphs are oftentimes observed both in experiment and simulations.
Properly quantifying the relative free energies of different possible crystal polymorphs and metastable structures is essential for informing how to design a system and associated conditions to preferentially form one structure, but this presents a difficult sampling problem given the presence of high free energy barriers in these systems.
Enhanced sampling methods that selectively bias a system along its relevant slow degree(s) of freedom associated with the transition(s) of interest,  i.e., collective variables (CVs), provide a means of efficiently mapping out these systems' free energy landscapes\cite{Bussi2020using,Henin2022enhanced,tuckerman_book,Giberti2015metadynamics}.
However, the identification of a set of CVs that properly capture the relevant degrees of freedom is essential for obtaining accurate, well-converged free energy estimates via these approaches.
Commonly used order parameters (OPs) like Steinhardt-Nelson parameters\cite{Steinhardt1983bond} and coordination numbers are effective CVs for biasing the sampling of simple crystalline systems\cite{Auer2001prediction,Trudu2006freezing,Filion2010crystal,Badin2021nucleating} but are less efficient when more exotic, non-close packed structures are involved as they can fail to resolve the different states of interest.
More complex OPs can better resolve the relevant (meta)stable states of a system, but the identification of these can be challenging given the many-body character of nucleation and crystallization processes.
A diverse library of both analytical\cite{Steinhardt1983bond,Lechner2008accurate,Mickel2013shortcomings,Piaggi2019calculation,Fijan2026quantifying,Russo2016crystal,Larsen2016robust,Teich2019identity} and data-driven\cite{Dietrich2024machine,Meraz2024simulating,Rogal2019neural,Karmakar2021collective} OPs have been developed to better enable enhanced sampling simulations to study colloidal crystal systems.
However, many analytical approaches can be too computationally demanding to serve as CVs for on-the-fly biased enhanced sampling simulations that require repeated differentiation of the OPs with respect to particle positions\cite{Fijan2026quantifying,Zhao2026hybrid}.
On the other hand, data-driven approaches have often required additional reweighting procedures to correct for the machine learning (ML) models' mispredictions\cite{Dietrich2024machine,Meraz2024simulating,Dietrich2026committor}.

Here we have developed OPs that distinguish the relevant crystal structures under a given set of PACS conditions.
Our OPs generalize a previously developed local environment similarity OP \cite{Piaggi2019calculation} to render them rotationally invariant.
However, this requires performing an expensive alignment of each particle's local environment with respect to the target crystal structure(s) that makes them computationally infeasible to use for biased enhanced sampling simulation to determine the system's free energy landscape.
To circumvent this computational bottleneck, we have trained SE(3) equivariant transformer networks to predict the values of these OPs.
The computational acceleration afforded by these ML proxy models enabled us to employ them as CVs to drive  metadynamics\cite{Laio2002escaping,Bussi2020using} simulations of ionic colloidal systems.
We used an active learning protocol to iteratively construct our training dataset by sampling new candidate configurations via metadynamics simulations and demonstrate that the ML model trained on the final dataset is accurate with respect to the analytical CV even over the course of lengthy metadynamics simulations.
We showcase the utility of our approach by computing and comparing free energy landscapes for ionic colloidal systems that can form both or either CsCl-like\cite{cscl} and Th$_3$P$_4$-like\cite{th3p4} crystals, highlighting how tuning the charges of the colloidal particles stabilizes one crystal structure over the other. 

\section{Methods and Computational Details}
\label{sec:methods}

\subsection{Polymer-attenuated Coulombic self-assembly (PACS)}
\label{sec:methods:pacs}

We have previously shown that binary PACS systems consisting of positive and negative ionic colloidal particles can be designed to self-assemble into a wide variety of colloidal crystals by changing particle size ratio, surface potential/potential ratio, and salt concentration \cite{Zang2025direct,vanKesteren2026light,vanKesteren2026structure}.
In Ref.~\citenum{Zang2025direct}, we observed particularly rich behavior for the size ratio 1:0.81, whereby a number of identified and unidentified crystals are observed under a custom dialysis setup. 
In simulation and experiment we observe a wide range of conditions where structures equivalent to the atomic crystals of CsCl, i.e., binary simple cubic crystals, and Th$_3$P$_4$ form (Fig.~\ref{fig:ref-structs}); moreover, we have observed heterogeneous nucleation of a previously unobserved structure possessing a stoichiometric ratio of 3:4 between the larger and smaller particles on a charged substrate \cite{Zang2025direct}.

While these structures appear in standard MD simulations using the coarse-grained potential described below, we were not previously able to identify for some conditions whether either Th$_3$P$_4$ or CsCl was thermodynamically favored or only kinetically favored. Moreover, for the majority of conditions where Th$_3$P$_4$ formed, it took substantially longer than the simpler CsCl structure as nucleation tended to proceed through an amorphous droplet phase before eventually crystallizing after O($10^9$) MD steps. 
Hence, for conditions where only aggregates are observed, we require a more advanced approach to determine thermodynamic stability.
Here, we demonstrate that this is possible using a generalized environment similarity CV. 

Like our previous studies\cite{Hueckel2020ionic,Zang2024structures,Zang2025direct,vanKesteren2026light}, we performed molecular dynamics (MD) simulations with a coarse-grained representation of the system such that each colloid is represented as a single particle interacting with other colloids via a pairwise potential that is a sum of screened electrostatic interactions due to innate surface charge and purely repulsive interactions due to overlap of polymer brushes on the surface.
We have developed a flexible simulation software framework, PACSim, that facilitates performing simulations with this model \cite{Hollmer2026pacsim}; below, we briefly describe the model for completeness but we refer the interested reader to Ref.~\citenum{Hollmer2026pacsim} for full details.

We model the electrostatic potential energy ($V_e$) between pairs of colloids $i$ and $j$ according to Derjaguin-Landau-Verwey-Overbeek theory as \cite{Hunter2001foundations,Hollmer2026pacsim},
\begin{equation}
    \label{eqn:coulomb}
    \frac{V_\mathrm{e}(h_{ij})}{k_{B}T} = 2\pi\epsilon a_{ij}\psi_{i}\psi_{j} e^{-h_{ij}/\lambda_{d}},
\end{equation}
where $h_{ij}$ is the distance separating the surfaces of the two colloids, $a_{ij}$ is the harmonic mean of the particles' radii following the Derjaguin approximation\cite{Hunter2001foundations}, $\lambda_{d}$ is the Debye screening length, $k_B$ is the Boltzmann constant, $T$ is the temperature, $\epsilon$ is the solvent permittivity, and $\psi_{i}$ and $\psi_{j}$ are the surface potentials of the respective colloids. 

We model the repulsive potential energy ($V_p$) associated with interactions between polymer brushes according to the Alexander-de Gennes polymer brush model\cite{Alexander1977polymer,Degennes1985,likos2000colloidal,Hollmer2026pacsim} as follows,
\begin{equation}
\begin{split}
    \frac{V_\mathrm{p}(h_{ij})}{k_{B}T} =& \frac{16\pi a_{ij}L^{2}\sigma^{3/2}}{35}\left[ 28\left(\left( \frac{2L}{h_{ij}}\right)^{1/4}-1\right)\right.\\
    &+ \left.\frac{20}{11}\left(1-\left( \frac{h_{ij}}{2L} \right)^{1/4}\right) + 12\left(\frac{h_{ij}}{2L}-1\right) \right],
\end{split}
\end{equation}
where $L$ is the polymer brush length and $\sigma$ is the surface brush density.

Due to the relative expense of computing these CVs, we employed a smaller system than used for earlier studies. Here, MD simulations consisted of 256 colloids in an equimolar mixture between positive and negative particles. Diameters of the positive and negative particles were 170~nm and 210~nm, respectively, in a periodic box with side lengths of 1800~nm.
We modeled our colloids interacting in a dilute salt solution, taking the solvent permittivity to be $\epsilon=80$ for that of water and the Debye screening length $\lambda=5.42$~nm.
For the polymer brush parameters, we set the brush length to $L=10.0$~nm and the surface brush density to be 0.09~nm$^{-2}$.
To test the effect of changing the surface potential ratio, we varied the surface potentials of the negative and positive colloids setting them to be either -40:+53, -53:+40, -70:+30~mV.
Given the product form of the prefactor in the electrostatic attraction in Eq.~\ref{eqn:coulomb}, these charges result in approximately identical pairwise attractions, while changing the scale of the like-charge repulsion.

We performed Langevin dynamics simulations, thermostatted to 300~K using a drag coefficient of 0.1~ps$^{-1}$ and a timestep of 0.05~ps.
Configuration files to reproduce these simulations are available as described in the Data Availability section.

\subsection{Local environment order parameter}
\label{sec:methods:ops}

For a given positive particle $i$, we defined the following OP to quantify the overlap of its local environment $\chi_i$ with respect to a reference local environment $\chi_\mathrm{P}$ (also centered on a positive particle),
\begin{equation}
\label{eqn:op_p}
\begin{split}
    O_\mathrm{P}(i\in P) =& \max_{\textbf{R}_{\mathrm{P}},\textbf{R}_{\mathrm{N}}} \frac{1}{2n_\mathrm{P}}\sum_{j\in\chi_i}^{n_\mathrm{P}}\sum_{k\in\chi_\mathrm{P}}^{n_\mathrm{P}} \exp{\left(-\frac{|\vec{r}_{ij}-\textbf{R}_\mathrm{P}\vec{r}_{k}|^{2}}{4\sigma_\mathrm{P}^{2}}\right)}\\
    &+ \frac{1}{2n_\mathrm{N}}\sum_{j\in\chi_i}^{n_\mathrm{N}}\sum_{k\in\chi_\mathrm{P}}^{n_\mathrm{N}} \exp{\left(-\frac{|\vec{r}_{ij}-\textbf{R}_\mathrm{N}\vec{r}_{k}|^{2}}{4\sigma_\mathrm{N}^{2}}\right)},
\end{split}
\end{equation}
where $n_\mathrm{P}$ is the number of positive particles in the local environment not including the central particle, $n_\mathrm{N}$ is the number of negative particles in the local environment, $\vec{r}_{ij}$ is the displacement between particle $j$ in environment $\chi_i$ and the centered particle $i$, $\vec{r}_{k}$ is the position of particle $k$ in the reference environment $\chi_\mathrm{P}$ where the central particle is located at the origin, $\sigma$ dictates the breadth of a particle's density, and $\textbf{R}_\mathrm{P}$ and $\textbf{R}_\mathrm{N}$ are rotation matrices that are separately determined for the positive P and negative N neighbors. Note that the indices $j$ and $k$ in the first set of summation terms run over the positive particles in the local environments $\chi_{i}$ and $\chi_{P}$, respectively, but they run over the negative particles in the second set of summation terms.

We similarly defined an OP for a given negative particle $i$ with respect to a reference local environment $\chi_\mathrm{N}$ to be,
\begin{equation}
\label{eqn:op_n}
\begin{split}
    O_\mathrm{N}(i\in N) =& \max_{\textbf{R}_{\mathrm{P}},\textbf{R}_{\mathrm{N}}} \frac{1}{2n_\mathrm{P}}\sum_{j\in\chi_i}^{n_\mathrm{P}}\sum_{k\in\chi_N}^{n_\mathrm{P}} \exp{\left(-\frac{|\vec{r}_{ij}-\textbf{R}_\mathrm{P}\vec{r}_{k}|^{2}}{4\sigma_\mathrm{P}^{2}}\right)}\\
    &+ \frac{1}{2n_\mathrm{N}}\sum_{j\in\chi_i}^{n_\mathrm{N}}\sum_{k\in\chi_\mathrm{N}}^{n_\mathrm{N}} \exp{\left(-\frac{|\vec{r}_{ij}-\textbf{R}_\mathrm{N}\vec{r}_{k}|^{2}}{4\sigma_\mathrm{N}^{2}}\right)}.
\end{split}
\end{equation}

As stated earlier, this is an adaptation of the environment similarity CV from Ref.~\citenum{Piaggi2019calculation} which is available in the PLUMED open source sampling library \cite{plumed-consort-2019} and illustrated in the PLUMED-tutorial example 22.012 \cite{tribello2025tutorials}.

\begin{figure}
    \centering
    \includegraphics[width=1\linewidth]{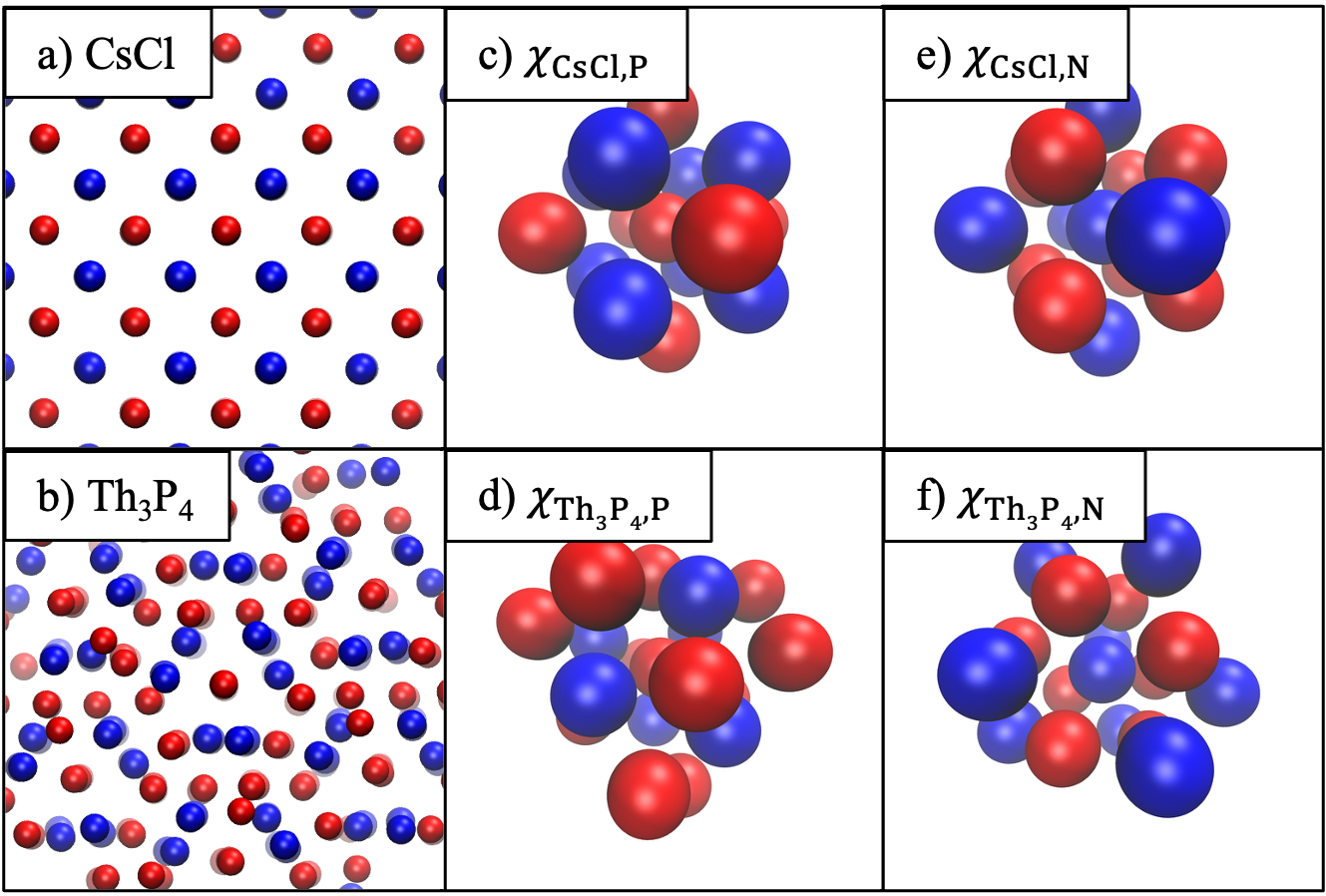}
    \caption{Energy-minimized structures for CsCl-like (a) and Th$_3$P$_4$-like (b) colloidal crystals viewed from an orthographic perspective and with the particle diameters set to 50\% of their actual values. The system consists of both positive (red) and negative (blue) colloidal particles with diameters of 170 and 210~nm, respectively. Reference local environments $\chi_\mathrm{CsCl,P}$ and $\chi_\mathrm{Th_{3}P_{4},P}$ around the positive particles are depicted in (c) and (d) while the reference local environments $\chi_\mathrm{CsCl,N}$ and $\chi_\mathrm{Th_{3}P_{4},N}$ around the negative particles are depicted in (e) and (f).}
    \label{fig:ref-structs}
\end{figure}

For CsCl-like crystals, the reference local environments for the positive ($X_\mathrm{CsCl,P}$) and negative ($X_\mathrm{CsCl,N}$) particles are that of a binary simple cubic crystal where there are 8 equidistant oppositely charged nearest neighbors (Figures~\ref{fig:ref-structs}c and ~\ref{fig:ref-structs}e).
Both the calculations of $O_\mathrm{CsCl,P}$ and $O_\mathrm{CsCl,N}$ use an overlap parameter $\sigma_\mathrm{N}=\sigma_{P}=23$~nm, but the former is computed with $n_\mathrm{N}=8$ and $n_\mathrm{P}=6$ whereas the calculation of the latter uses $n_\mathrm{N}=6$ and $n_\mathrm{P}=8$.
For Th$_{3}$P$_{4}$-like crystals, the reference local environment for the smaller positive particles ($X_\mathrm{Th_3P_4,P}$) consists of $n_\mathrm{N}=6$ and $n_\mathrm{P}=11$ neighbors and overlap parameters of $\sigma_\mathrm{N}=23$~nm and $\sigma_\mathrm{P}=45$~nm whereas the reference local environment for the negative particles ($X_\mathrm{Th_3P_4,N}$) consists of $n_\mathrm{N}=8$ and $n_\mathrm{P}=8$ neighbors and overlap parameters of $\sigma_\mathrm{N}=64$~nm and $\sigma_\mathrm{P}=32$~nm.

Note that the calculation of these OPs for each particle $i$'s local environment involves a rotational alignment to determine the maximum overlap of $\chi_i$ with the corresponding reference environment.
We employed the Kabsch algorithm\cite{Kabsch1976solution} to determine the rotation matrices $\mathrm{\textbf{R}}$ that give the maximal alignment between the set of particles in particle $i$'s local environment and those of the reference environments.

\subsection{Metadynamics and collective variables}
\label{sec:methods:metadynamics}

\begin{figure*}
    \centering
    \includegraphics[width=1\textwidth]{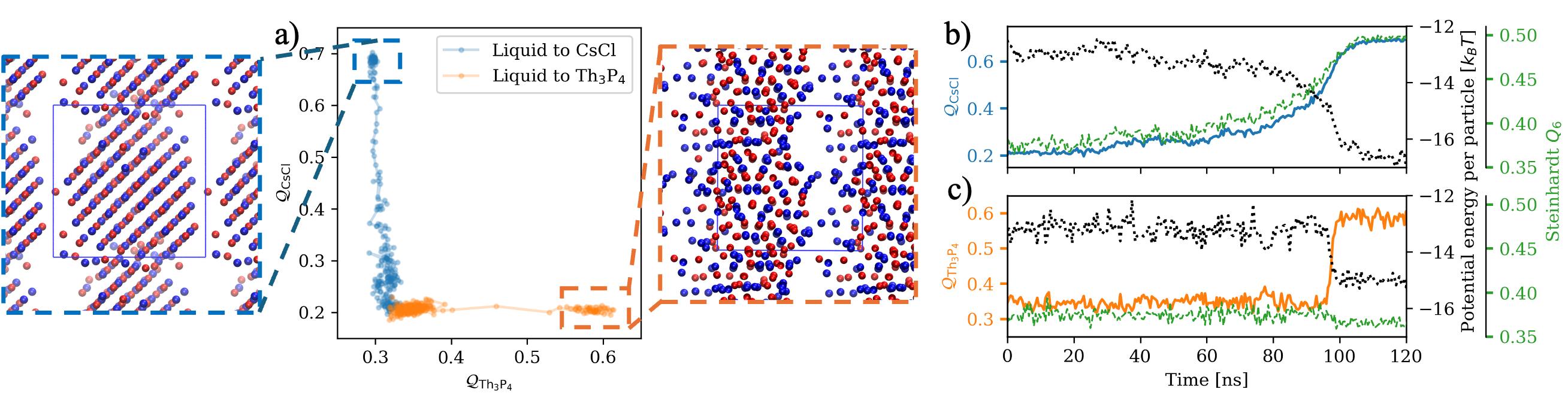}
    \caption{a) Our CVs, $Q_\mathrm{CsCl}$ and $Q_\mathrm{Th_3P_4}$, over the course of an unbiased MD simulation starting from a disordered liquid that crystallized into CsCl (blue) where surface potentials of the negative and positive particles were -40 and +53 mV, respectively, and another simulation that resulted in a Th$_3$P$_4$ crystal when using surface potentials of -53 and +40 mV. The CVs resolve the crystalline phases from each other and from the disordered phases. b+c) Time traces of the system's $Q_\mathrm{CsCl}$ and $Q_\mathrm{Th_3P_4}$ (blue and orange) track with the potential energy (black) during unbiased MD simulations along with the Steinhardt-Nelson $Q_6$ OP for comparison (green).}
    \label{fig:unbiased}
\end{figure*}

We performed (untempered) metadynamics simulations to efficiently sample the relevant (meta)stable states for our ionic colloidal crystal system and compute their relative free energies.
This involves augmenting our sampling of a system's canonical distribution with a history-dependent bias potential where Gaussian hills are deposited in the space of a defined set of CVs.
After observing a number of transitions between metastable states of the system, we can estimate the free energy of the system by taking the negative of the summed bias potential \cite{Laio2002escaping}.
Here, we employed untempered metadynamics because we were most interested in evaluating whether our CVs allowed us to explore the space of structures, and significant bias is required to transition over very high energy barriers. 

Having defined local OPs that should distinguish our target crystal structures of interest, we applied a metadynamics bias on two global CVs, one that serves as a measure of how CsCl-like the system is,
\begin{align}
\label{eqn:cv_cscl}
    Q_\mathrm{CsCl} = \frac{1}{2N_\mathrm{P}}\sum_{i=1}^{N_\mathrm{P}} \bar{O}_\mathrm{CsCl,P}(i) + \frac{1}{2N_\mathrm{N}}\sum_{j=1}^{N_\mathrm{N}} \bar{O}_\mathrm{CsCl,N}(j),
\end{align}
and another that serves as a measure of how Th$_3$P$_4$-like the system is,
\begin{align}
\label{eqn:cv_th3p4}
    Q_\mathrm{Th_3P_4} = \frac{1}{2N_\mathrm{P}}\sum_{i=1}^{N_\mathrm{P}} \bar{O}_\mathrm{Th_3P_4,P}(i) + \frac{1}{2N_\mathrm{N}}\sum_{j=1}^{N_\mathrm{N}} \bar{O}_\mathrm{Th_3P_4,N}(j),
\end{align}
where the indices $i$ and $j$ run over the set of positive and negative colloidal particles in our systems with totals of $N_\mathrm{P}$ and $N_\mathrm{N}$, respectively. For calculating the CV of each crystal structure $\alpha$ we used their respective neighbor-averaged OPs,

\begin{equation}
\begin{split}
    \bar{O}_{\mathrm{\alpha,P}}(i) =& \frac{1}{2(1+n_{\mathrm{P}})}\left(O_{\alpha,\mathrm{P}}(i) + \sum_{j\in \chi_{i}}^{n_{\mathrm{P}}} O_{\alpha,\mathrm{P}}(j)\right)\\
    &+ \frac{1}{2n_{\mathrm{N}}}\sum_{k\in\chi_{i}}^{n_{\mathrm{N}}} O_{\alpha,\mathrm{N}}(k),
\end{split}
\end{equation}

\begin{equation}
\begin{split}
    \bar{O}_{\mathrm{\alpha,N}}(i) =& \frac{1}{2(1+n_{\mathrm{N}})}\left(O_{\alpha,\mathrm{N}}(i) + \sum_{j\in \chi_{i}}^{n_{\mathrm{N}}} O_{\alpha,\mathrm{N}}(j)\right)\\
    &+ \frac{1}{2n_{\mathrm{P}}}\sum_{k\in\chi_{i}}^{n_{\mathrm{P}}} O_{\alpha,\mathrm{P}}(k),
\end{split}
\end{equation}
in a fashion similar to how neighbor-averaged local Steinhardt order parameters are often used\cite{Lechner2008accurate}.

The Gaussian hills added to the potential over the course of our metadynamics simulations were deposited every 100~ps (2000 MD steps).
We set the heights of the deposited Gaussian energy biases to be 1~kJ/mol and their standard deviations to be 0.025 for both $Q_\mathrm{CsCl}$ and $Q_\mathrm{Th_3P_4}$.

\subsection{Machine learning model and dataset construction}
\label{sec:methods:ml}

Since the evaluation of the CVs in Section~\ref{sec:methods:metadynamics} and their gradients with respect to particle positions would computationally bottleneck our metadynamics simulations, we have developed ML models to serve as more efficient-to-evaluate proxies for both $Q_\mathrm{CsCl}$ and $Q_\mathrm{Th_3P_4}$.
We employed the equivariant transformer (ET) network architecture\cite{tholke2022torchmd-net}, which has been used extensively for modeling ML potentials for various molecular systems\cite{Chen2025machine,Munoz2025grotthuss}, as implemented in the TorchMD-Net package\cite{pelaez2024torchmdnet}.
However, instead of using it as a map from atomic positions to the system's potential energy, here we developed ET models that serve as a map between the positions of the colloidal particles and our CVs.
Our ET models for both $Q_\mathrm{CsCl}$ and $Q_\mathrm{Th_3P_4}$ employed a learnable 32-dimensional embedding for describing the local environment around a given colloid, with a lower radial cutoff of 180~nm and an upper radial cutoff of 380~nm, constructed from 64 radial basis functions.
These initial particle embeddings were refined with two update layers that use a modified attention mechanism consisting of four attention heads\cite{Vaswani2017attention}.
In all, each of our ET models consisted of 50300 trainable parameters.

For each CV $\alpha$, we minimized a loss function for a given configuration of $N_N$ negative and $N_P$ positive colloids that combines the square error of both the global CVs and per-particle local environment similarity OPs,
\begin{equation}
\begin{split}
    L_\alpha =& \lambda_{Q} \left( \widehat{Q}_{\alpha} - Q_{\alpha}\right)^{2}\\
    &+ \frac{\lambda_{O}}{N_\mathrm{P}}\sum_{i}^{N_\mathrm{P}} \left(\widehat{O}_{\alpha,\mathrm{P}}(i)-O_{\alpha,\mathrm{P}}(i)\right)^{2} \\
    &+ \frac{\lambda_{O}}{N_\mathrm{N}}\sum_{i}^{N_\mathrm{N}} \left(\widehat{O}_{\alpha,\mathrm{N}}(i)-O_{\alpha,\mathrm{N}}(i)\right)^{2},
\end{split}
\end{equation}
where the terms with carets indicate the corresponding ML model predictions, and the $\lambda$ coefficients are used to weight the relative contribution between global and local losses.
For training our models, we set $\lambda_{Q}=0.1$ and $\lambda_{O}=100$ to more heavily weight the training on the local environment order parameters.
We nevertheless still employ a non-zero weight for the global term so as to reduce any sort of systematic bias in the local predictions.
To optimize the weights of the ML models, we used the Adam\cite{kingma2015adam} optimizer with a learning rate of 0.001, $\beta_1=0.9$, and $\beta_2=0.999$.

Our dataset contains a total of 11355 periodic configurations each consisting of 256 colloidal particles and we employed a 90-10 training-validation split to train our ML models.
We constructed the dataset via an active learning protocol\cite{krogh1994neural} where we iteratively sampled new candidate training set configurations by performing metadynamics with our ML proxy CVs, post-processed the sampled trajectory to identify configurations that were mispredicted, and added those failure modes to the training set for subsequent iterations.
The initial dataset consisted of 250 configurations sampled from each of two unbiased MD simulations depicted in Figure~\ref{fig:unbiased}: (1) one simulation where the system was initialized in a disordered liquid state and self-assembled into a CsCl-like crystal by the end and (2) another simulation that self-assembled into a Th$_3$P$_4$-like crystal.
We then performed several active learning iterations employing a variety of different ML proxy CVs that evaluated how similar a given configuration's structure is with the target structures' radial distribution functions (RDF) and angularly-resolved RDFs.
Although we found these CVs to be insufficient for comprehensive configuration sampling via metadynamics, during this process we added a diverse set of 9713 configurations.
In particular, we were never able to sample Th$_3$P$_4$-like crystals with these earlier CVs.
The remaining 1142 configurations were sampled via a second round of active learning where we now employed our local environment OPs as defined in Section~\ref{sec:methods:ops} as our CVs.

\section{Results and Discussion}
\label{sec:results_discussion}

\begin{figure}
    \centering
    \includegraphics[width=\linewidth]{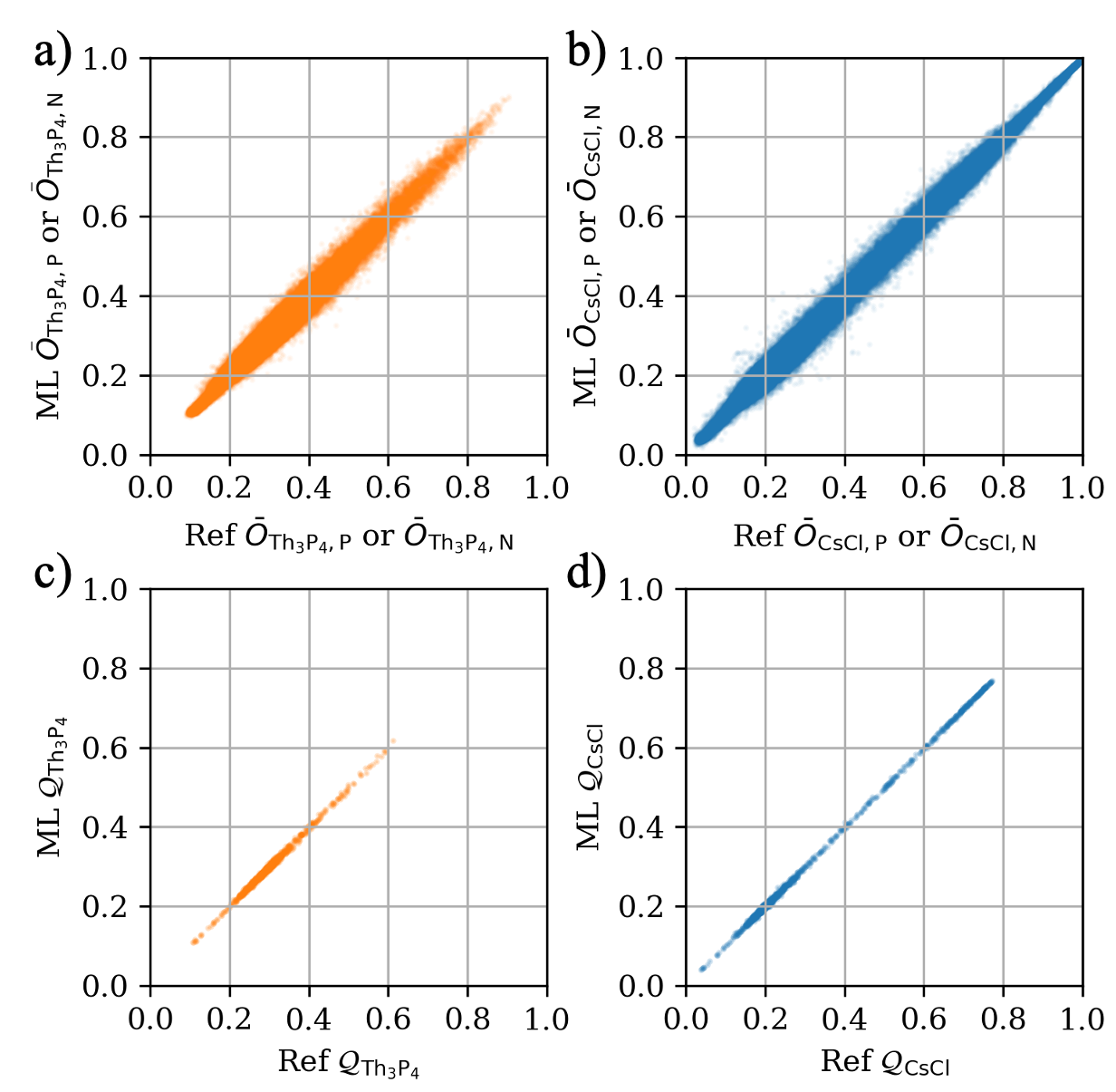}
    \caption{The accuracy of predictions obtained from our ML models trained on reference particle-specific order parameters (OPs) gauging how similar their local environment is compared to that of Th$_3$P$_4$-like (a) and CsCl-like (b) crystals. Our CVs (c+d) are the configuration average of our OPs and the values obtained from our ML models are well correlated with the reference values. The correlation plots show the ML predicted values as compared to the reference values as evaluated on our 1135 configuration validation set.}
    \label{fig:ml-accuracy}
\end{figure}

\begin{figure*}
    \centering
    \includegraphics[width=0.9\textwidth]{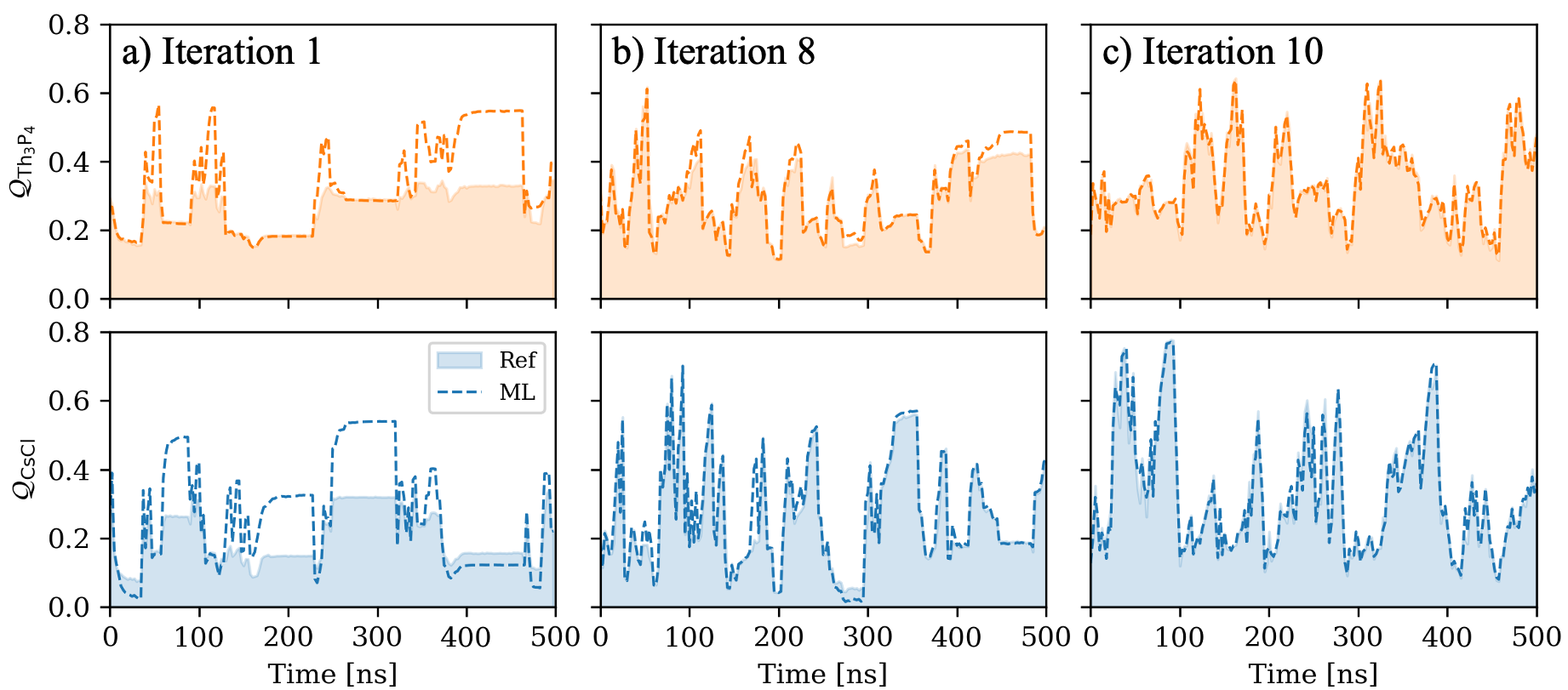}
    \caption{Time traces of $Q_\mathrm{CsCl}$ and $Q_\mathrm{Th_3P_4}$ over the course of a metadynamics simulation performed at iterations 1, 8, and 10 of our active learning protocol for constructing the dataset where the ML predicted CVs used to drive the metadynamics simulation are shown as dashed lines and the reference values for the CVs are represented by the shaded area. (a) Our ML model of both $Q_\mathrm{CsCl}$ and $Q_\mathrm{Th_3P_4}$ trained to the initial dataset displays large errors. By iteration 8 our ML model for $Q_\mathrm{CsCl}$ is accurate but our model for $Q_\mathrm{Th_3P_4}$ still displays errors at higher values, however both models accurately predict the CVs by iteration 10.}
    \label{fig:active-learning-iters}
\end{figure*}

To estimate free energies of the relevant colloidal crystal structures via metadynamics, our CVs should accurately resolve the system's different metastable and transition states.
Figure~\ref{fig:unbiased}a tracks both the CVs for how CsCl-like ($Q_\mathrm{CsCl}$) and Th$_3$P$_4$-like ($Q_\mathrm{Th_3P_4}$) our 256 particle periodic system is over the course of unbiased MD simulations where we initialize the system with a disordered liquid-state initial configuration and observed it self-assemble into a CsCl-like (blue line) or Th$_3$P$_4$-like (orange) crystal.
Both unbiased MD simulations were performed with the settings specified in Section~\ref{sec:methods:pacs} with the surface potentials of the negative and positive colloids set to -40 and +53~mV for the simulation that produced the CsCl crystal and -53 and +40~mV for the other. By construction (Equations~\ref{eqn:op_p})-\ref{eqn:op_n}), the domain for each of our local environment order parameters runs from 0, where there is no overlap of neighboring particles with that of the reference structure, to 1, where the displacements of the neighboring particles match those of the reference structure when optimally aligned.
Hence, our CVs (Equations~\ref{eqn:cv_cscl}-\ref{eqn:cv_th3p4}), which are averages of the local environment OPs of each particle in the system, will also run from 0 to 1 with greater values indicating that more particles in the system have structured themselves into the particular reference crystal structure.
The time trace of $Q_\mathrm{CsCl}$ over the course of the unbiased MD simulation presented in Figure~\ref{fig:unbiased}b shows that it properly resolves the initial disordered liquid state from the fully structured CsCl-like crystal at $Q_\mathrm{CsCl}\approx 0.7$.
Similarly, Figure~\ref{fig:unbiased}c shows that $Q_\mathrm{Th_3P_4}$ well-resolves the initial liquid state from when the system ultimately self-assembles into a Th$_3$P$_4$-like crystal beyond 100~ns at $Q_{Th_3P_4}\approx 0.6$.
Both $Q_\mathrm{CsCl}$ and $Q_\mathrm{Th_3P_4}$ also capture intermediate transition states as the system nucleates and grows the crystal seeing as their respective time trace tracks with that of the system's potential energy as the crystals self-assemble (dashed lines in Figures~\ref{fig:unbiased}b-c).
Although decreases in the potential energy like those in Figures~\ref{fig:unbiased}b-c are strong indicators of the system crystallizing, the potential energy would be a poor CV given the multitude of degenerate structures that present with identical energies.
On the other hand, we see from Figure~\ref{fig:unbiased}a that regions in the 2D space spanned by $Q_\mathrm{CsCl}$ and Q$_\mathrm{Th_3P_4}$ that are representative of our two crystal structures of interest, as well as the associated transition states, are well-separated from each other and from the disordered state.
Although in principle these CVs with their ability to resolve the relevant states of our colloidal system appear to be well-suited for performing metadynamics simulations, in practice they would be infeasible to employ because of the computational challenge associated with computing their gradients with respect to particle positions given the rotational alignment needed to determine the maximum alignment of each particle's local environment with the reference environments (Equations~\ref{eqn:op_p}-\ref{eqn:op_n}). 

To enable the use of $Q_\mathrm{CsCl}$ and $Q_\mathrm{Th_3P_4}$ for metadynamics simulations, we have trained computationally efficient ML proxy models to a dataset of the local environment OPs (Section~\ref{sec:methods:ml}).
Figure~\ref{fig:ml-accuracy} highlights the accuracy of our ML proxy models as evaluated on a validation set of 1135 periodic configurations each consisting of 256 colloidal particles.
The correlation plots in Figures~\ref{fig:ml-accuracy}a-d show good agreement between the ML predicted global and local order parameters.
The performance of our ML models over a diverse validation set of local environments spanning the entire range of both $O_{CsCl}$ and $O_{Th_3P_4}$, and their configuration averages that serve as our CVs (Figures~\ref{fig:ml-accuracy}c-d), gives us confidence that we can use our ML models as proxies for our $Q_\mathrm{CsCl}$ and $Q_\mathrm{Th_3P_4}$.
In addition to checking prediction accuracies on a select validation set of configurations, we also over the course of our active learning procedure for iteratively constructing the dataset (Section~\ref{sec:methods:ml}) sampled metadynamics trajectories employing our best-of-yet ML proxy CVs and post-processed the trajectories to compute the reference CV values for identifying problematic regions of configuration space that could be added to the dataset to improve our models.
Figure~\ref{fig:active-learning-iters}a shows the ML predicted values of $Q_\mathrm{CsCl}$ and $Q_\mathrm{Th_3P_4}$ used to drive the metadynamics simulation for iteration 1 of our active learning protocol.
Over the course of this initial 500~ns metadynamics simulation, which was conducted for a system where the colloids either had +46 or -46~mV surface potentials, we observed that the ML predicted $Q_\mathrm{CsCl}$ and $Q_\mathrm{Th_3P_4}$ values deviated significantly from the reference values at the higher end of each of the CVs' domains.
Consequently, our metadynamics simulation for iteration 1, which was initialized in the disordered liquid phase, failed to sample configurations where the system crystallized into CsCl or Th$_3$P$_4$.
After refining our dataset iteratively via our active learning protocol (Section~\ref{sec:methods:ml}), our ML proxy CVs at iteration 8 were robust enough to enable our metadynamics simulations to start sampling both crystal structures.
However, Figure~\ref{fig:active-learning-iters}b shows that our ML proxy for $Q_\mathrm{Th_3P_4}$ still occasionally made erroneous over-predictions (i.e., between 400 and 500~ns).
It was relatively more challenging to develop the ML proxy CV for Th$_3$P$_4$ because its crystal structure presents with more complex local environments than that of the interlaced simple cubic structure of CsCl (Figure~\ref{fig:ref-structs}).
Nevertheless, by iteration 10 both of our ML proxy CVs were consistently accurate over the course of the metadynamics simulations (Figure~\ref{fig:active-learning-iters}c).

\begin{figure*}
    \centering
    \includegraphics[width=0.9\textwidth]{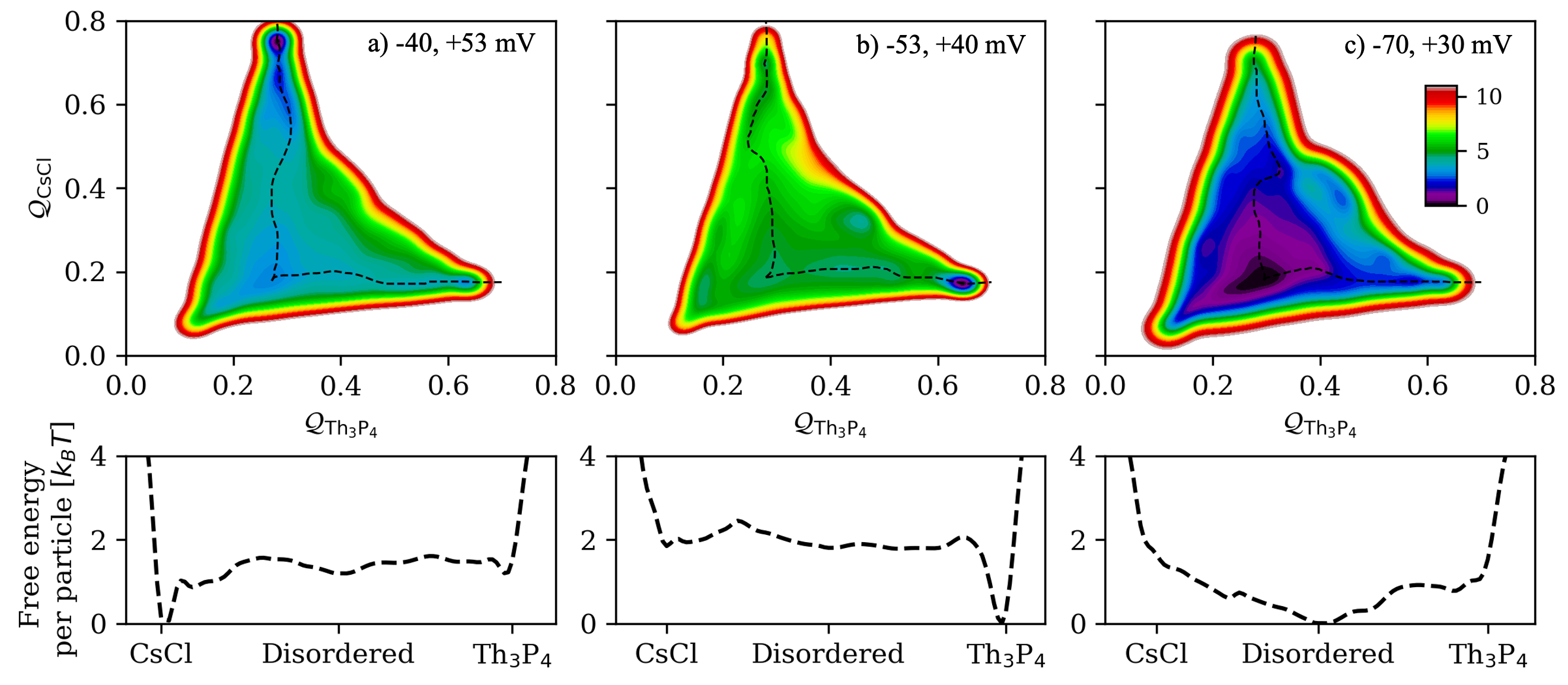}
    \caption{2D free energy surfaces obtained from our metadynamics simulations employing our ML models of $Q_\mathrm{CsCl}$ and $Q_\mathrm{Th_3P_4}$ when varying the surface potentials of the negative and positive particles showing how CsCl and Th$_3$P$_4$ crystals can be thermodynamically favored under different conditions. Dashed lines show minimum free energy paths along the surface, with 1D free energies shown in the lower panels. In all 3 cases the negative particle has a diameter of 210~nm and the positive particle has a diameter of 170~nm. (a) When the surface potentials for negative and positive particles are -40 and +53~mV (-40,+53~mV), respectively, then CsCl-like crystals are favored. (b) When we flip things to -53,+40~mV we get a strong relative stabilization of Th$_3$P$_4$-like crystals. (c) If we continue redistributing the surface potentials to -70,+30~mV we destabilize both crystals with respect to the disordered phases.}
    \label{fig:surface-pots}
\end{figure*}

We leveraged our accurate and more efficient ML proxy CVs to understand how the free energy landscape of our ionic colloidal systems varies as we modulate the surface potentials of the positive and negative colloidal particles while holding all other parameters fixed.
For consistency, we selected surface potential combinations that would keep the Coulombic interaction between pairs of positive and negative colloids similar by fixing their product $\psi_{i}\psi_{j}$ to approximately -2100~mV$^2$.
Figure~\ref{fig:surface-pots} presents the results of our free energy calculations where we employed our ML proxies for $Q_\mathrm{CsCl}$ and $Q_\mathrm{Th_3P_4}$ within metadynamics simulations.
Here, we find that when we set the surface potentials of the larger negatively charged colloids to -40~mV and the smaller positive colloids to +53~mV, CsCl-like crystals are strongly stabilized with respect to Th$_3$P$_4$-like crystals and the disordered phase (Figure~\ref{fig:surface-pots}a).
By shifting more and more of the charge onto the larger negative particles, CsCl-like structures became more destabilized while Th$_3$P$_4$-like structures became more stabilized.
More specifically, when we flipped the surface potentials to -53,+40~mV Th$_3$P$_4$-like crystals present with considerably lower free energies (Figure~\ref{fig:surface-pots}b).
We can rationalize this change by considering how they affect the pairwise Coulombic interactions (Equation~\ref{eqn:coulomb}).

Hence, the differences in relative free energies we observe when tuning the surface potentials are attributable to changes in the repulsive Coulombic interactions between particles of the same type.
More specifically, the larger negative particles are packed closer together on average when they adopt a CsCl-like structure as compared to when they structure like Th$_3$P$_4$ (241~nm vs. 282~nm).
When the magnitude of the surface potentials on the negative particles is smaller (e.g., Figure ~\ref{fig:surface-pots}a), neighboring negative particles are able to pack closer together into CsCl-like structures without incurring too much of an energetic penalty.
On the other hand, when the magnitude of the surface potentials on the negative particles is greater, the expected distances between neighboring negative particles in a CsCl-like arrangement lie further up the repulsive wall ($>1~k_BT$) while negative particles for Th$_3$P$_4$ are separated further apart and unaffected.
As a consequence, as we see from Figure ~\ref{fig:surface-pots}b for this case, Th$_3$P$_4$-like structures have a lower free energy than CsCl-like structures.
However if we further increase the magnitude of the surface potentials on the negative particles (Figure ~\ref{fig:surface-pots}c) then the repulsive interactions at larger distances increase to the point where Th$_3$P$_4$-like structures are also destabilized and instead disordered condensates will have lower free energies. 
This is consistent with our recent experimental study where we demonstrated that particular crystal types can be selected by using surfactants to independently modulate the surface potentials of the positive and negative colloids; although that study used different sized particles, we independently came to the same conclusion as to the effects of like-like repulsion on tipping the balance between CsCl and Th$_3$P$_4$ \cite{vanKesteren2026structure}.
This shows how our use of local environment CVs to compute relative free energies can predict the results of experimentally-accessible routes to directing the self-assembly of specific colloidal crystal structures.

\section{Conclusions}
\label{sec:conclusion}

Metadynamics simulations can be useful for characterizing the free energy landscape of colloidal crystal systems, but require identifying a set of CVs that serve as low dimensional representations of the system capable of resolving the different metastable structures and associated transition states while being efficient to differentiate with respect to particle positions.
Here we presented an approach that uses CVs measuring the similarity of local particle environments with those of relevant perfect crystal structures, and developed ML proxies for these CVs so that we could employ them efficiently for metadynamics simulations.
We showed that one can construct datasets for training these ML models via an iterative active learning procedure such that they serve as accurate proxies.
We demonstrated the utility of our approach by using the trained ML proxies for determining the free energy surface landscapes of ionic colloidal systems that tend to form CsCl-like and/or Th$_3$P$_4$-like crystals.
Our results showed that one can vary the surface potentials of the positive and negative colloidal particles in these systems to preferentially stabilize one of these crystal structures, and the trends we obtain from our metadynamics simulations employing our ML proxy CVs are qualitatively consistent with experiments conducted on similar systems.

The approach we developed here enables the use of CVs that would otherwise be too computationally expensive or infeasible to evaluate the gradients for on-the-fly biased enhanced sampling simulations of colloidal crystal systems.
Note that the calculation of our local environment order parameters necessitates that we have a predefined set of reference environments.
If different reference environments need to be used, either because the new system has different relevant crystal structures and/or particle sizes, then our ML proxy models need to be retrained accordingly.
However, as we demonstrated here, training these system-specific ML proxy models can be achieved in a data efficient manner.
The development of these models could be further sped up by leveraging transfer learning between similar systems (e.g., from one set of colloid diameters to another). 
We believe this approach will generalize and be useful for calculating relative free energies of other colloidal crystal systems going forward.


\section{Data Availability}
The datasets used to train our ML models, codes used to train those models and run our simulations, scripts to compute order parameters, and inputs for performing the simulations are available at https://github.com/hocky-research-group/Chen-ML-proxyCVs.

\begin{acknowledgments}
This research was supported by the US Army Research Office under award number W911NF-26-2-A149 to SS and GMH. GMH also acknowledges support of a fellowship from the Alfred P. Sloan foundation. 
M.S.C. was supported as a fellow of the Simons Center for Computational Physical Chemistry at
NYU (SCCPC, Simons Foundation Grant MPS-T-MPS-00839534, MET). M.S.C. gratefully acknowledges startup funds from the University of Kentucky. This work was supported in part through the NYU IT High Performance Computing resources, services, and staff expertise.
\end{acknowledgments}

\section*{References}
\bibliography{library}

\end{document}